\documentstyle[preprint,aps,epsfig]{revtex}

\draft

\begin{document}
\title{Nucleon-nucleon cross sections in neutron-rich matter and isospin
transport in heavy-ion reactions at intermediate energies}
\author{{\bf Bao-An Li$^{1}$ and Lie-Wen Chen$^{2,3}$}}
\address{$^1$ Department of Chemistry and Physics, P.O. Box 419,\\
Arkansas State University, State University, Arkansas 72467-0419, USA\\
$^2$ Institute of Theoretical Physics, 
Shanghai Jiao Tong University, Shanghai 200240, China\\
$^{3}$ Center of Theoretical Nuclear Physics, National Laboratory of Heavy
Ion Accelerator, Lanzhou 730000, China}
\maketitle

\begin{abstract}
Nucleon-nucleon (NN) cross sections are evaluated in neutron-rich matter
using a scaling model according to nucleon effective masses. It is found that 
the in-medium NN cross sections are not only reduced but also have a different
isospin dependence compared with the free-space ones. Because of the
neutron-proton effective mass splitting the difference between nn and pp
scattering cross sections increases with the increasing isospin asymmetry of
the medium. Within the transport model IBUU04, the in-medium NN cross
sections are found to influence significantly the isospin transport in
heavy-ion reactions. With the in-medium NN cross sections, a symmetry energy
of $E_{sym}(\rho )\approx 31.6(\rho /\rho _{0})^{0.69}$ was found most
acceptable compared with both the MSU isospin diffusion data and the
presently acceptable neutron-skin thickness in $^{208}$Pb. The isospin
dependent part $K_{asy}(\rho _{0})$ of isobaric nuclear incompressibility
was further narrowed down to $-500\pm 50$ MeV. The possibility of
determining simultaneously the in-medium NN cross sections and the symmetry
energy was also studied. The proton transverse flow, or even better the
combined transverse flow of neutrons and protons, can be used as a probe of
the in-medium NN cross sections without much hindrance from the
uncertainties of the symmetry energy. \newline
\end{abstract}


\bigskip

\newpage

\section{Introduction}

The isospin dependence of in-medium nuclear effective interactions is
important for understanding not only novel properties of exotic nuclei near
drip lines but also many interesting questions in astrophysics \cite%
{lat01,steiner,riatheory}. Especially, it determines the symmetry energy $E_{%
\text{sym}}(\rho )$ term in the equation of state (EOS) of isospin
asymmetric nuclear matter. The density-dependent symmetry energy itself is
still poorly known but very important for both nuclear physics and
astrophysics. Heavy-ion reactions induced by neutron-rich nuclei provide a
unique opportunity to explore the isospin dependence of in-medium nuclear
effective interactions, especially the symmetry energy, in a broad range of
density. This is because the isospin degree of freedom plays an important
role in heavy-ion collisions through both the nuclear EOS and the
nucleon-nucleon (NN) scatterings \cite{ireview,ibook}. In particular, the
transport of isospin asymmetry between two colliding nuclei is expected to
depend on both the symmetry potential and the in-medium NN cross sections.
For instance, the drifting contribution to the isospin transport in a nearly
equilibrium system is proportional to the product of the mean relaxation
time $\tau _{np}$ and the isospin asymmetric force $F_{np}$ \cite{shi}.
While the $\tau _{np}$ is inversely proportional to the neutron-proton (np)
scattering cross section $\sigma _{np}$ \cite{shi}, the $F_{np}$ is directly
related to the gradient of the symmetry potential. On the other hand, the
collisional contribution to the isospin transport in non-equilibrium system
is generally expected to be proportional to the np scattering cross section.
Thus the isospin transport in heavy-ion reactions depends on both the
long-range and the short-range parts of the isospin-dependent in-medium
nuclear effective interactions, namely, the symmetry potential and the
in-medium np scatterings cross sections. The former relates directly to the
density dependence of the symmetry energy $E_{\text{sym}}(\rho )$. To
extract information about $E_{\text{sym}}(\rho )$ has been one of the major
goals of heavy-ion reactions induced by neutron-rich nuclei \cite%
{ireview,ibook}. Among the potential probes proposed so far, the isospin
transport has been found very useful for investigating the $E_{\text{sym}%
}(\rho )$ \cite{shi,betty04,chen05}. More specifically, using the isospin
and momentum dependent transport model IBUU04 \cite{ibuu04}, a symmetry
energy of the form $E_{\text{sym}}(\rho )\approx 32(\rho /\rho _{0})^{1.1}$
was extracted recently from the MSU data on isospin transport \cite{chen05}.
This conclusion, however, was drawn based on transport model calculations
using the experimental free-space NN cross sections. In this work, we
examine effects of the in-medium NN cross sections on isospin transport in
heavy-ion collisions at intermediate energies. The in-medium NN cross
sections are calculated within a scaling model according to the nucleon
effective masses consistent with the nuclear mean field used in the
transport model. It is found that the density dependence of symmetry energy
extracted from the isospin transport data is altered significantly. The
possibility of determining both the in-medium NN cross sections and the
symmetry energy corresponding to the same underlying nuclear effective
interactions is also discussed.

\section{Momentum dependence of the mean field and nucleon effective masses
in neutron-rich matter}

The single nucleon potential is one of the most important inputs to all
transport models for nuclear reactions. Both the isovector (symmetry
potential) and isoscalar parts of this potential should be momentum
dependent due to the non-locality of strong interactions and the Pauli
exchange effects in many-fermion systems. In the IBUU04 transport model \cite%
{ibuu04}, we use a single nucleon potential derived within the Hartree-Fock
approach using a modified Gogny effective interaction (MDI) \cite{das03},
i.e.,
\begin{eqnarray}
U(\rho ,\delta ,\vec{p},\tau ,x) &=&A_{u}(x)\frac{\rho _{\tau ^{\prime }}}{%
\rho _{0}}+A_{l}(x)\frac{\rho _{\tau }}{\rho _{0}}  \nonumber \\
&+&B(\frac{\rho }{\rho _{0}})^{\sigma }(1-x\delta ^{2})-8\tau x\frac{B}{%
\sigma +1}\frac{\rho ^{\sigma -1}}{\rho _{0}^{\sigma }}\delta \rho _{\tau
^{\prime }}  \nonumber \\
&+&\frac{2C_{\tau ,\tau }}{\rho _{0}}\int d^{3}p^{\prime }\frac{f_{\tau }(%
\vec{r},\vec{p}^{\prime })}{1+(\vec{p}-\vec{p}^{\prime })^{2}/\Lambda ^{2}}
\nonumber \\
&+&\frac{2C_{\tau ,\tau ^{\prime }}}{\rho _{0}}\int d^{3}p^{\prime }\frac{%
f_{\tau ^{\prime }}(\vec{r},\vec{p}^{\prime })}{1+(\vec{p}-\vec{p}^{\prime
})^{2}/\Lambda ^{2}}.  \label{mdi}
\end{eqnarray}

\begin{figure}[tbh]
\includegraphics{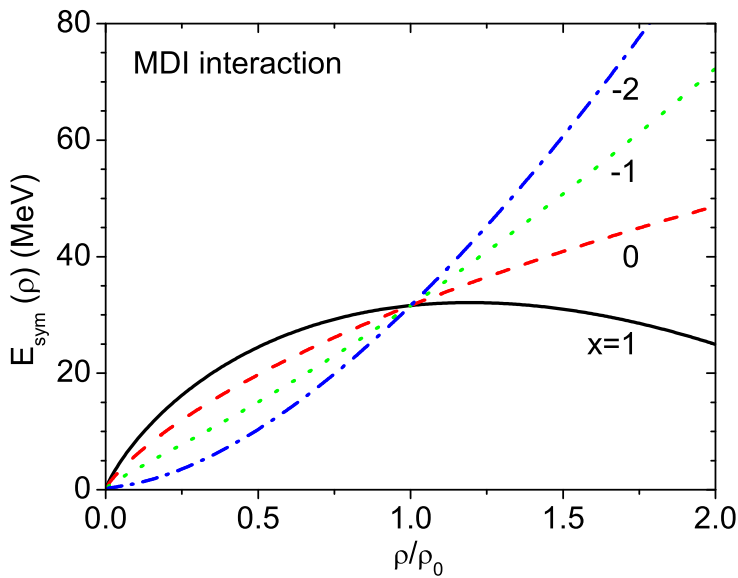} 
\vspace*{1cm}
\caption{(Color online) Symmetry energy as a function of density for the MDI
interaction with $x=1,0,-1$ and $-2$. Taken from ref.\protect\cite{chen05}.}
\label{figure1}
\end{figure}
Here $\delta =(\rho _{n}-\rho _{p})/\rho $ is the isospin asymmetry of the
nuclear medium. In the above $\tau =1/2$ ($-1/2$) for neutrons (protons) and
$\tau \neq \tau ^{\prime }$; $\sigma =4/3$; $f_{\tau }(\vec{r},\vec{p})$ is
the phase space distribution function at coordinate $\vec{r}$ and momentum $%
\vec{p}$. The parameters $A_{u}(x),A_{l}(x),B,C_{\tau ,\tau },C_{\tau ,\tau
^{\prime }}$ and $\Lambda $ were obtained by fitting the momentum-dependence
of the $U(\rho ,\delta ,\vec{p},\tau ,x)$ to that predicted by the Gogny
Hartree-Fock and/or the Brueckner-Hartree-Fock (BHF) calculations \cite%
{bombaci}, the saturation properties of symmetric nuclear matter and the
symmetry energy of about $30$ MeV at normal nuclear matter density $\rho
_{0}=0.16$ fm$^{-3}$ \cite{das03}. The incompressibility $K_{0}$ of
symmetric nuclear matter at $\rho _{0}$ is set to be $211$ MeV consistent
with the latest conclusion from studying giant resonances \cite%
{k0data,pie04,colo04}. The parameters $A_{u}(x)$ and $A_{l}(x)$ depend on
the $x$ parameter according to
\begin{equation}
A_{u}(x)=-95.98-x\frac{2B}{\sigma +1},~~~~A_{l}(x)=-120.57+x\frac{2B}{\sigma
+1}.
\end{equation}%
The parameter $x$ can be adjusted to mimic predictions on the density
dependence of symmetry energy $E_{\text{sym}}(\rho )$ by microscopic and/or
phenomenological many-body theories. Shown in Fig.\ 1 is the density
dependence of the symmetry energy for $x=-2$, $-1$, $0$ and $1$.
The last two terms in Eq. (\ref{mdi}) contain the momentum-dependence of the
single-particle potential. The momentum dependence of the symmetry potential
stems from the different interaction strength parameters $C_{\tau ,\tau
^{\prime }}$ and $C_{\tau ,\tau }$ for a nucleon of isospin $\tau $
interacting, respectively, with unlike and like nucleons in the background
fields. More specifically, we use $C_{unlike}=-103.4$ MeV and $C_{like}=-11.7
$ MeV. With these parameters, the isoscalar potential estimated from $%
(U_{neutron}+U_{proton})/2$ agrees reasonably well with predictions from the
variational many-body theory \cite{wiringa}, the more advanced BHF approach
\cite{zuo} including three-body forces and the Dirac-Brueckner-Hartree-Fock
(DBHF) calculations\cite{sam05} in broad ranges of density and momentum.

What is particularly interesting and important for nuclear reactions induced
by neutron-rich nuclei is the isovector (symmetry) potential. The strength
of this potential can be estimated very accurately from $%
(U_{neutron}-U_{proton})/2\delta $ \cite{ibuu04}. In Fig.\ 2 the strength of
the symmetry potential for four $x$ parameters is displayed as a function of
momentum and density. Here we have only plotted the symmetry potential at
sub-saturation densities most relevant to heavy-ion reactions studies at
intermediate energies. It is noticed that the momentum dependence of the
symmetry potential is independent of the parameter $x$. This is because the $%
x$ appears only in the density dependent part of the single nucleon
potential of Eq.\ (\ref{mdi}) by construction. Systematic analyses of a
large number of nucleon-nucleus scattering experiments and (p,n) charge
exchange reactions at beam energies below about 100 MeV indicate undoubtedly
that the symmetry potential at $\rho _{0}$, i.e., the Lane potential,
decreases approximately linearly with increasing beam energy $E_{kin}$. The
data can be well described by using $U_{Lane}=a-bE_{kin}$ where $a\simeq
22-34$ MeV and $b\simeq 0.1-0.2$ \cite{data1,data2,data3,data4}. One should
comment that although the uncertainties in both the parameters $a$ and $b$
are large, the decreasing feature of the Lane potential with increasing beam
energy is very certain. This provides a stringent constraint on the symmetry
potential. The potential in Eq. (\ref{mdi}) at $\rho _{0}$ satisfies this
requirement very well as seen in Fig. 2. What is very uncertain but most
interesting is the momentum-dependent symmetry potential at abnormal
densities. The experimental determination of both the density and momentum
dependence of the symmetry potential is thus desired. Heavy-ion reactions
provide a unique tool in terrestrial laboratories to explore the symmetry
potential at abnormal densities with varying momenta.
\begin{figure}[tbh]
\vspace*{0.5cm} 
\includegraphics[height=0.55%
\textheight]
{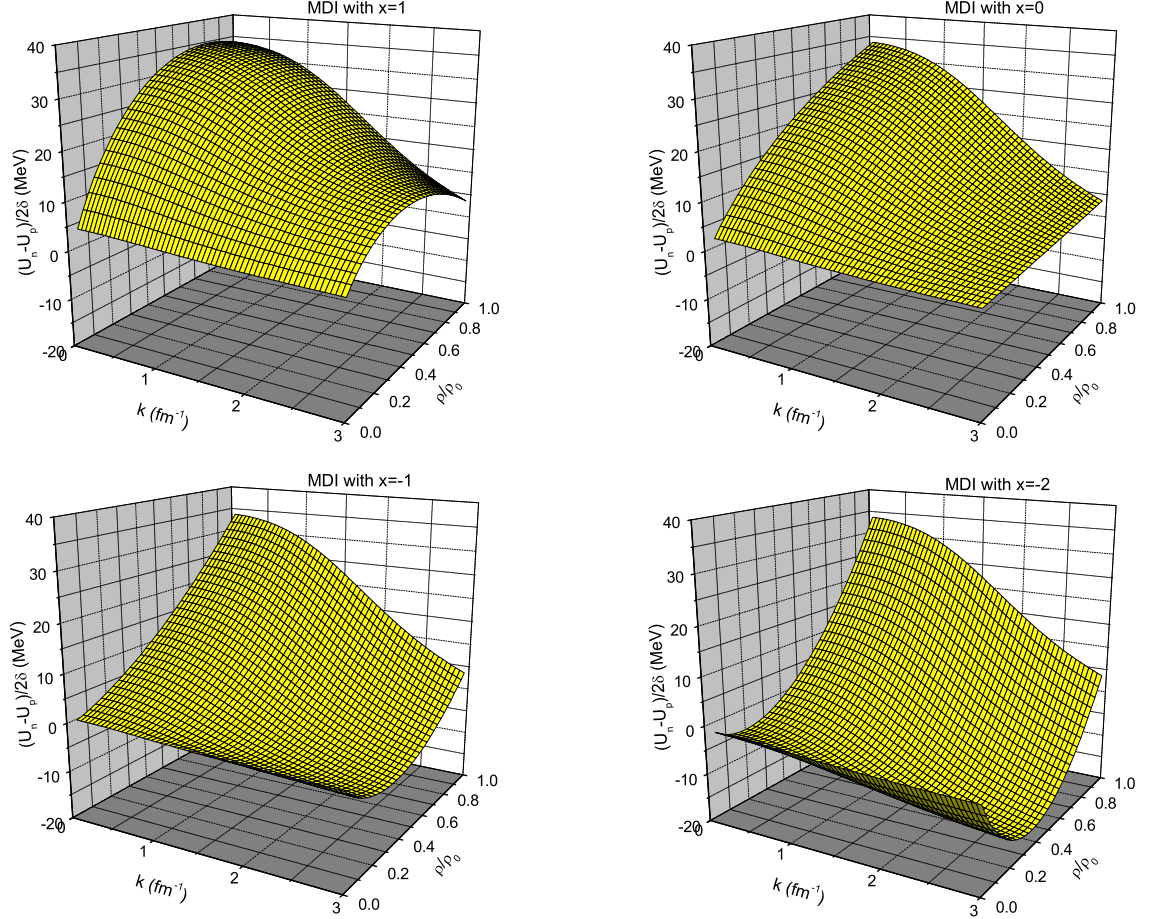} 
\vspace*{1cm}
\caption{(Color online) Symmetry potential as a function of momentum and
density for MDI interactions with $x=1,0,-1$ and $-2$. }
\label{figure2}
\end{figure}
One characteristic feature of the momentum dependence of the symmetry
potential is the different effective masses for neutrons and protons in
isospin asymmetric nuclear matter, i.e.,
\begin{equation}
\frac{m_{\tau }^{\ast }}{m_{\tau }}=\left\{ 1+\frac{m_{\tau }}{p}\frac{%
dU_{\tau }}{dp}\right\} .  \label{emass}
\end{equation}%
By definition, the effective mass normally depends on the density and
isospin asymmetry of the medium as well as the momentum of the nucleon.
Conventionally, however, the effective mass at the Fermi momentum $p_{\tau
}=p_{f}({\tau })$ is most frequently used to characterize the momentum
dependence of the nuclear potential. In our calculations of the in-medium NN
cross sections, we use the general definition of Eq. (\ref{emass}). With the
potential in Eq. (\ref{mdi}), since the momentum-dependent part of the
nuclear potential is independent of the parameter $x$, the nucleon effective
masses are independent of the $x$ parameter too.
\begin{figure}[tbh]
\vspace*{0.4cm} \includegraphics[height=0.45%
\textheight,angle=-90]{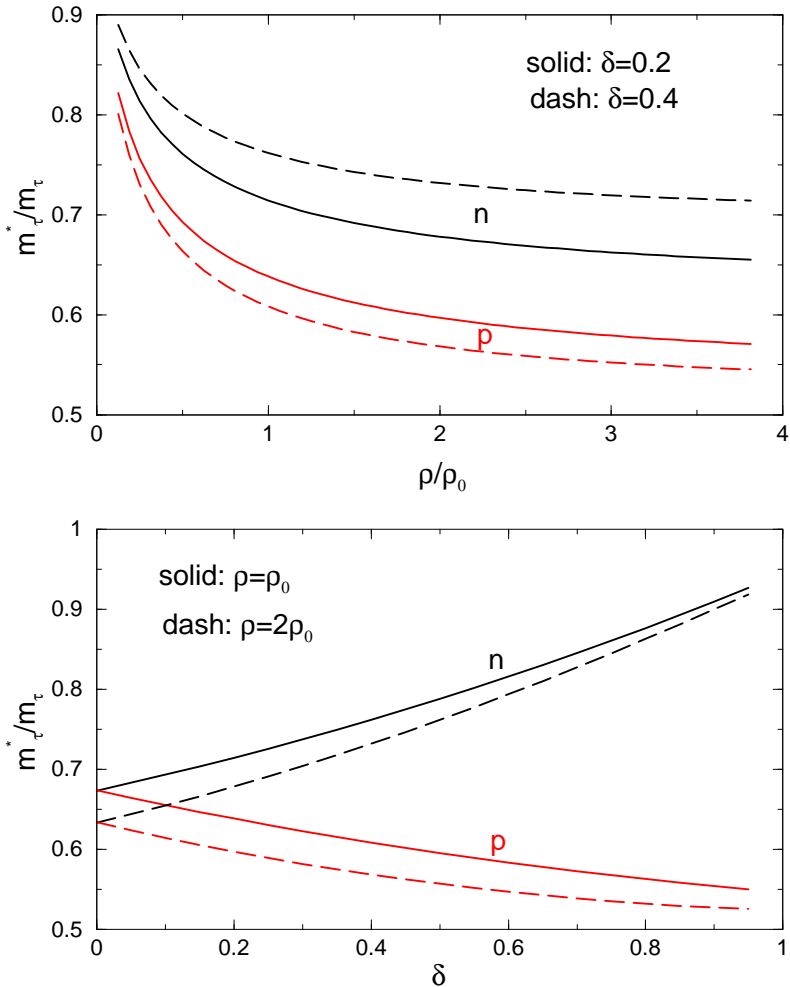} 
\vspace*{1cm}
\caption{{\protect\small (Color online) Nucleon effective masses at the
respective Fermi surface in asymmetric matter as a function of density
(upper window) and isospin asymmetry (lower window).}}
\label{figure3}
\end{figure}

Shown in Fig. 3 are the effective masses of neutrons and protons at their
respective Fermi surfaces as a function of density (upper window) and
isospin asymmetry (lower window). It is seen that the effective mass of
neutrons is higher than that of protons and the splitting between them
increases with both the density and isospin asymmetry of the medium. We
notice here that the momentum dependence of the symmetry potential and the
associated neutron-proton effective mass splitting is still highly
controversial within different approaches and/or using different nuclear
effective interactions \cite{ditoro04,li04,beh}. Being phenomenological and
non-relativistic in nature the neutron-proton effective mass splitting in
the present study is consistent with predictions of all non-relativistic
microscopic models, see, e.g., \cite{bombaci,zuo,sjo}, and the
non-relativistic limit of microscopic relativistic many-body theories, see,
e.g., \cite{sam05,ma,fuchs1}. Recent transport model studies indicate that
the neutron/proton ratio at high transverse momenta and/or rapidities is a
potentially useful probe of the neutron-proton effective mass splitting in
neutron-rich matter \cite{ibuu04,ditoro05}. In this work we explore effects
of the neutron-proton effective mass splitting on in-medium NN cross
sections in neutron-rich matter. Applications of these in-medium NN cross
sections in heavy-ion reactions may be useful for finding other probes of
the nucleon effective masses in neutron-rich matter.

\section{Nucleon-nucleon cross sections in neutron-rich matter}

While much attention has been given to finding experimental observables
constraining the symmetry energy, little effort has been made so far to
study the NN cross sections in isospin asymmetric nuclear matter. Most of
the existing work on the in-medium NN cross sections have concentrated on
their density dependence in isospin symmetric nuclear matter, see, e.g.,
\cite{neg,pan,gqli,schu,gale,gg,mk,gli,chen01,pawel}. Here we extend the
effective mass scaling model for in-medium NN cross sections \cite%
{neg,pan,gale} to isospin asymmetric matter. Both the incoming current in
the initial state and the level density of the final state in NN scatterings
depend on the effective masses of the colliding nucleons. Assuming all
matrix elements of the NN interactions are the same in free-space and in the
medium, the NN cross sections in the medium $\sigma _{NN}^{medium}$ are
expected to be reduced compared with their free-space values $\sigma
_{NN}^{free}$ by a factor
\begin{equation}
R_{medium}\equiv \sigma _{NN}^{medium}/\sigma _{NN}^{free}=(\mu _{NN}^{\ast
}/\mu _{NN})^{2},  \label{xmedium}
\end{equation}%
where the $\mu _{NN}$ and $\mu _{NN}^{\ast }$ are the reduced masses of the
colliding nucleon pairs in free-space and in the medium, respectively.

At relative momenta less than about 240 MeV/c and densities less than about $%
2\rho _{0}$, the scaling of $\sigma _{NN}^{medium}/\sigma _{NN}^{free}$ in
Eq. (\ref{xmedium}) was recently found to be consistent with calculations
based on the DBHF theory \cite{fr}. This finding lends a strong support to
the effective mass scaling model of the in-medium NN cross sections in the
limited density and momentum ranges. In this work, we apply the scaling
model to elastic NN scatterings in heavy-ion reactions at beam energies up
to about the pion production threshold. At higher energies, inelastic
reaction channels become important and in-medium effects on these channels
have been a subject of much interest, see, e.g., refs. \cite%
{bert88,li96,cai05}. For the inelastic channels we keep using the
experimental free-space cross sections. This assumption has no effect on our
present study mainly at intermediate energies.

While the effective masses and the in-medium NN cross sections have to be
calculated dynamically in the evolving environment created during heavy-ion
reactions, it is instructive to examine the in-medium NN cross sections in
isospin asymmetric nuclear matter at zero temperature. In this situation the
integrals in Eq. (\ref{mdi}) can be analytically carried out. More
specifically \cite{das03,kuo},
\begin{eqnarray}
\int d^{3}p^{\prime }\frac{f_{\tau }(\vec{r},\vec{p}^{\prime })}{1+(\vec{p}-%
\vec{p}^{\prime })^{2}/\Lambda ^{2}} &=&\frac{2}{h^{3}}\pi \Lambda ^{3}\left[
\frac{p_{f}^{2}(\tau )+\Lambda ^{2}-p^{2}}{2p\Lambda }ln\frac{(p+p_{f}(\tau
))^{2}+\Lambda ^{2}}{(p-p_{f}(\tau ))^{2}+\Lambda ^{2}}\right.   \nonumber \\
&&\left. +\frac{2p_{f}(\tau )}{\Lambda }-2\{{\rm arctan}\frac{p+p_{f}(\tau )%
}{\Lambda }-{\rm arctan}\frac{p-p_{f}(\tau )}{\Lambda }\}\right] .
\end{eqnarray}%
The medium reduction factor $R_{medium}$ can thus also be obtained
analytically, albeit lengthy.
\begin{figure}[tbh]
\vspace*{-0.4cm} \includegraphics[height=0.55%
\textheight,angle=-90]{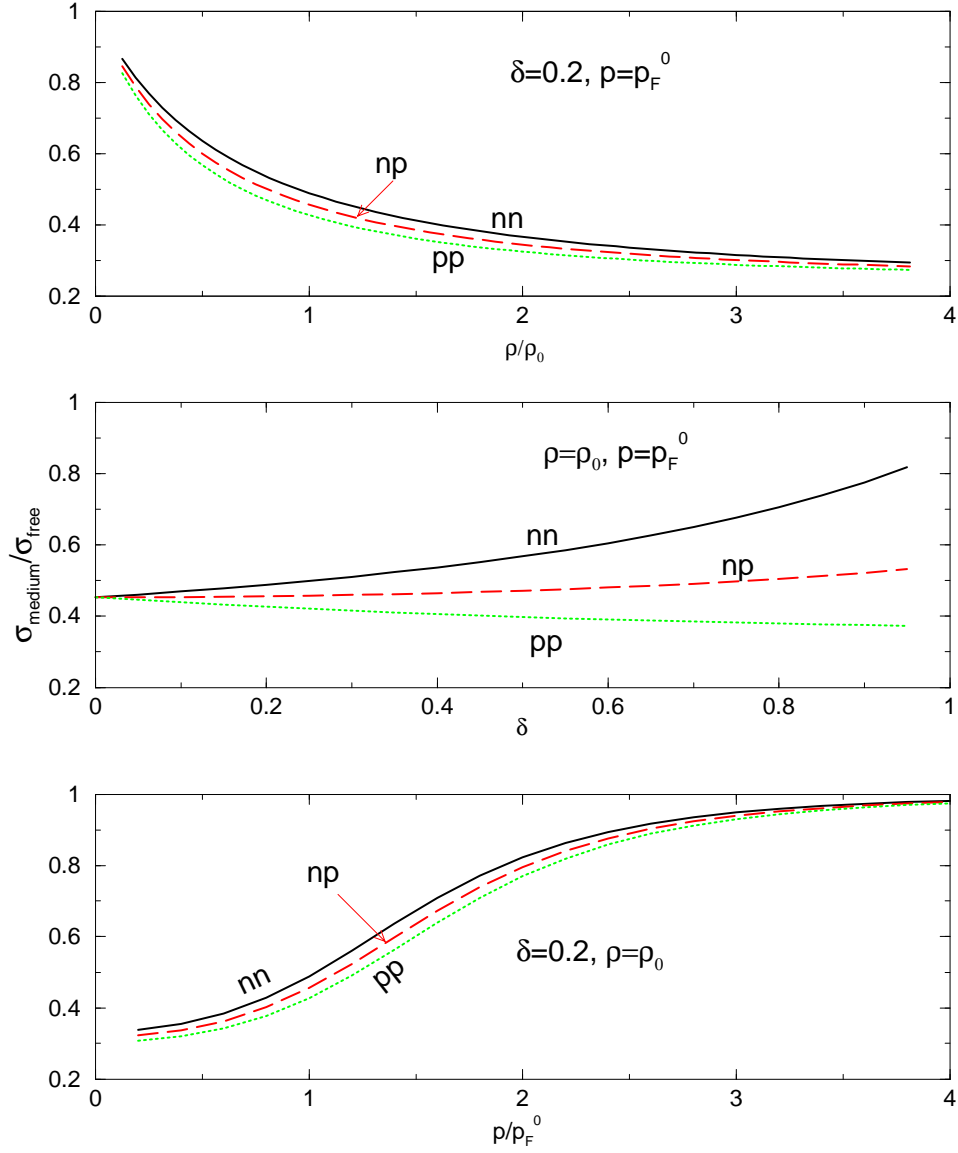} 
\vspace*{1cm}
\caption{{\protect\small (Color online) The reduction factor of the
in-medium nucleon-nucleon cross sections compared to their free-space values
as a function of density (top), isospin asymmetry (middle) and momentum
(bottom).}}
\label{figure4}
\end{figure}

As an illustration of a simplified case, we show in Fig. 4 the reduction
factor $R_{medium}$ for two colliding nucleons having the same momentum $p$.
The $R_{medium}$ factor is examined as a function of density (upper),
isospin asymmetry (middle) and the momentum (bottom). It is interesting to
note that the in-medium NN cross sections are not only reduced compared with
their free-space values, but the nn and pp cross sections are also split
while their free-space cross sections are the same. Moreover, the difference
between the nn and pp scattering cross sections grows in more asymmetric
matter. The higher in-medium cross sections for nn than for pp are
completely due to the positive neutron-proton effective mass splitting with
the effective interaction used. This feature may serve as a probe of the
neutron-proton effective mass splitting in neutron-rich matter. This
possibility will be studied in a future work. We also note that the
in-medium NN cross sections are also independent of the parameter $x$. They
are solely determined by the momentum dependence of the nuclear potential
used in the model.

\section{Dynamical generation of nucleon effective masses during heavy-ion
reactions}

The nucleon effective masses change dynamically in heavy-ion collisions. How
big are the nucleon effective masses and how much the NN cross sections are
modified compared with their free-space values in a typical heavy-ion
reaction at intermediate energies? To answer these questions, we show in
Fig.\ 5 the correlation between the average nucleon effective mass and the
average nucleon density (top), and the distribution of nucleon effective
masses (bottom) at the instant of $10$ fm/c in the reaction of $^{132}$Sn+$%
^{124}$Sn at a beam energy of $50$ MeV/A and an impact parameter of $5$ fm.
For this particular calculation $x=0$ is used, very similar results are
obtained using other values for the $x$ parameter. It is seen that the
nucleon effective masses decrease with increasing density. The maximum
density reached at the instant considered, i.e., $10$ fm/c, is about $%
1.4\rho /\rho _{0}$. Moreover, the neutron-proton effective mass
splitting is seen to increase slightly at supra-normal densities.
However, the increase is not much because the isospin asymmetry
normally decreases with increasing density, i.e., the isospin
fractionation (distillation), see, e.g.,
\cite{ireview,ibook,baran05}. These features are consistent with
our expectations discussed in previous sections. From the lower
window it is seen that the distribution of nucleon effective
masses picks at about $0.7$ GeV. It is also noticed that some
small number of nucleons obtain effective masses above their free
masses. This is also understandable. In principle, the slope of
the nucleon potential $du/dp$ in Eq. (\ref{emass}) can be negative
during heavy-ion reactions although it happens very rarely,
leading to the higher effective masses of some nucleons.
\begin{figure}[tbh]
\vspace*{0.4cm} \includegraphics[height=0.5%
\textheight,angle=-90]{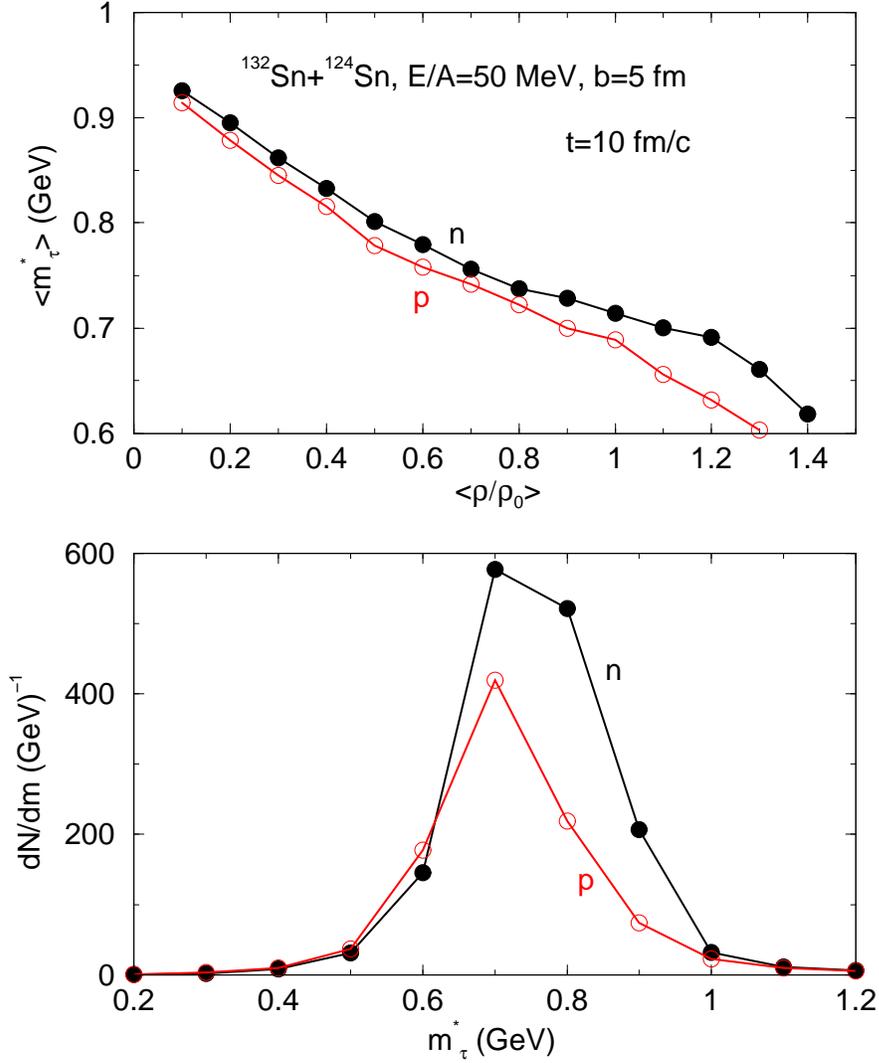} 
\vspace*{1cm}
\caption{{\protect\small (Color online) The correlation between the average
nucleon effective mass and the average nucleon density (top), and the
distribution of nucleon effective masses (bottom) in the reaction of }$^{132}
${\protect\small Sn+}$^{124}${\protect\small Sn at a beam energy of 50 MeV/A
and an impact parameter of 5fm.}}
\label{figure5}
\end{figure}
\begin{figure}[tbh]
\vspace*{0.4cm} \includegraphics[height=0.55%
\textheight,angle=-90]{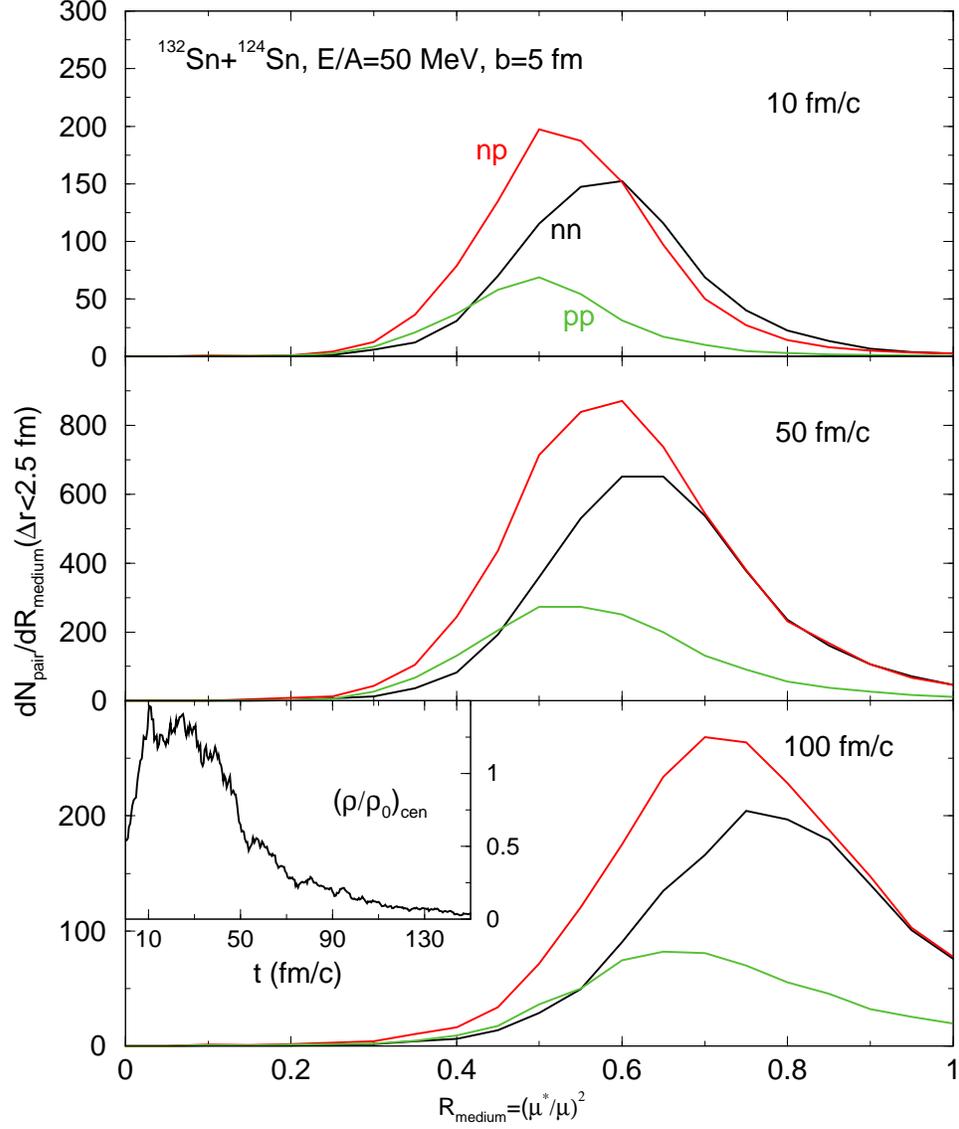} 
\vspace*{1cm}
\caption{{\protect\small (Color online) The distribution of the reduction
factor of in-medium NN cross sections in the reaction of }$^{132}$%
{\protect\small Sn+}$^{124}${\protect\small Sn at a beam energy of 50 MeV/A
and an impact parameter of 5fm at 10, 50 and 100 fm/c, respectively. The
insert is the evolution of the central density in the reaction.}}
\label{figure6}
\end{figure}
With the nucleon effective masses available we can now examine their effects
on nucleon-nucleon scatterings during heavy-ion reactions. Shown in Fig.\ 6
are the distributions of the reduction factor $R_{medium}$ in the reaction
of $^{132}$Sn+$^{124}$Sn at a beam energy of 50 MeV/A and an impact
parameter of $5$ fm at $10$, $50$ and $100$ fm/c, respectively. The insert
in the bottom window shows the evolution of the central density during the
reaction. The three instants represent the compression, expansion and
freeze-out stages of the reaction. The quantity $N_{pair}(\Delta r<2.5fm)$
is the number of nucleon pairs with spatial separations less than $2.5$ fm.
These are potential colliding nucleons whose scattering cross section will
be reduced by the factor $R_{medium}$, i.e., $\sigma
_{NN}^{medium}=R_{medium}\times \sigma _{NN}^{free}$. It is seen that on
average as much as $50\%$ reduction occurs for NN scatterings in the early
stage of the reaction. As the system expands the average density decreases,
the reduction factor $R_{medium}$ thus gradually shifts towards $1$ in the
later stage of the reaction.

\section{Effects of the in-medium NN cross sections on isospin transport in
heavy-ion reactions at intermediate energies}

The in-medium NN cross sections are expected to affect several aspects of
heavy-ion reactions. Previous studies have already found that the in-medium
NN cross sections influence both the degree and rate of isospin equilibrium
\cite{bass,lisherry1,chen97,liko98,mosel,liu}, see, e.g., \cite{lisherry2}
for a review. However, none of them is in the context of extracting the
symmetry energy. In this section we investigate how the symmetry energy
extracted from the isospin transport data might be altered.

\subsection{Isospin diffusion/transport}

Experimentally, one way of measuring quantitatively the amount of isospin
transport, between the projectile nucleus $A$ and the target nucleus $B$ is
to study the quantity $R_{i}$ defined as~\cite{rami}
\begin{equation}
R_{i}=\frac{2X^{A+B}-X^{A+A}-X^{B+B}}{X^{A+A}-X^{B+B}}  \label{Ri}
\end{equation}%
where $X$ is any isospin-sensitive observable. The $R_{i}$ is also known as
a useful tool of measuring quantitatively the degree of isospin diffusion in
heavy-ion reactions \cite{betty04}. Here we prefer to use the more general
term isospin transport although the two terminologies may have been used
interchangeably in the literature. Our main concern is that the term
diffusion is often associated with irreversible processes. While by
definition as given in Eq. (\ref{Ri}), and from analyzing its time evolution
in the following, it is seen that the $R_{i}$ does not always have to
decrease with time. In other words, there is no guarantee that the
particular observable $X$ is synchronized in all three reaction systems such
that $R_{i}$ always decreases with time. This is because the evolution and
freeze-out time of the observable $X$ may depend significantly on the system
size especially if the masses of $A$ and $B$ are rather different. By
construction, the value of $R_{i}$ is $1~(-1)$ for the symmetric $A+A~(B+B)$
reaction. If an isospin equilibrium is reached as a result of isospin
transport the value of $R_{i}$ is about zero. As mentioned earlier, the EOS
of symmetric nuclear matter still has some uncertainties associated with the
density and momentum dependence of the isoscalar potential. Fortunately,
these uncertainties and all others due to the isospin-independent
ingredients in the reaction dynamics can be largely canceled out because of
the special construction of $R_{i}$. The $R_{i}$ also has the advantage of
minimizing significantly effects of pre-equilibrium emissions \cite{betty04}%
.
\begin{figure}[th]
\includegraphics[scale=1.4]{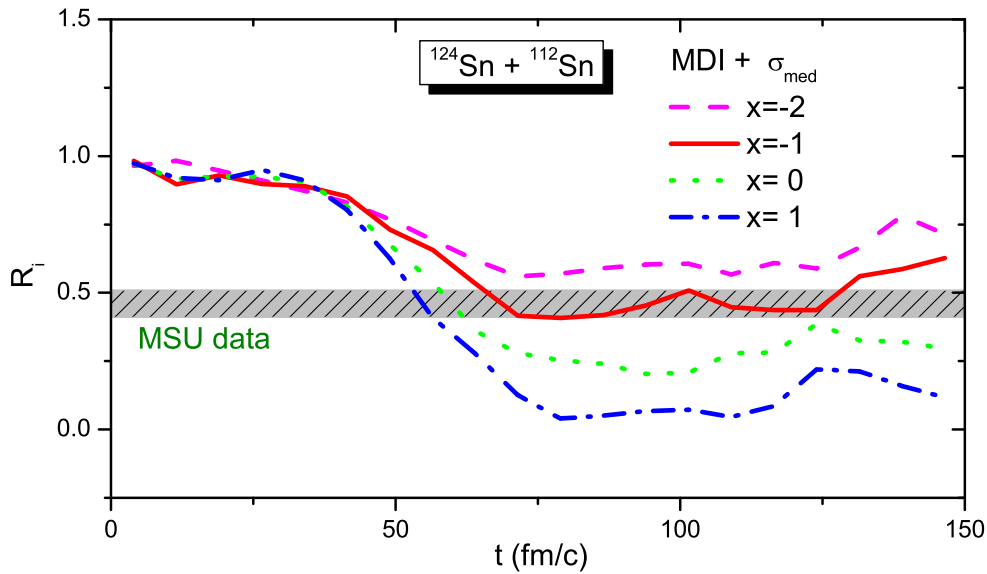}
\vspace{1.0 cm}
\caption{{\protect\small (Color online) The evolution of isospin diffusion $%
R_{i}$ using the four $x$ parameters and the in-medium nucleon-nucleon cross
sections.}}
\label{figure7}
\end{figure}
In the NSCL/MSU experiments with $A=$ $^{124}$Sn on $B=$ $^{112}$Sn at a
beam energy of $50$ MeV/nucleon and an impact parameter about $6$ fm, the
isospin asymmetry of the projectile-like residue was used as the isospin
tracer $X$ \cite{betty04}. The data was recently analyzed within the IBUU04
transport model using the free-space experimental NN cross sections.
Consistent with the experimental selection, in model analyses the average
isospin asymmetry $\left\langle \delta \right\rangle $ of the $^{124}$%
Sn-like residue was calculated from nucleons with local densities higher
than $\rho _{0}/20$ and velocities larger than $1/2$ the beam velocity in
the c.m. frame. Shown in Fig.\ 7 are the time evolutions of $R_{i}$
re-calculated using the in-medium NN cross sections and the four $x$
parameters. The data from MSU is indicated by the shaded band. It is seen
that the net isospin transport and the influence of the $x$ parameter show
up mainly in the expansion phase of the reaction after about $40$ fm/c. The
values of $R_{i}$ stabilize approximately after about $80$ fm/c. It is seen
that the values of $R_{i}$ with $x=-1$ and\ $0$ come close to the MSU data
in the late stage of the reaction.

For a more meaningful comparison with the experimental data, we have
calculated the time average of $R_{i}$ between $t=120$ fm/c and $150$ fm/c
as in Ref. \cite{chen05}. Shown in Fig.\ 8 is a comparison of the averaged
strength of isospin transport $1-R_{i}$ obtained with the free and in-medium
NN cross sections, respectively, as a function of the asymmetric part of the
isobaric incompressibility of nuclear matter at $\rho _{0}$ \cite%
{para,lopez88,baran05}
\begin{equation}
K_{\text{asy}}(\rho _{0})\equiv 9\rho _{0}^{2}\left( d^{2}E_{\text{sym}%
}/d\rho ^{2}\right) _{\rho _{0}}-18\rho _{0}\left( dE_{\text{sym}}/d\rho
\right) _{\rho _{0}}.
\end{equation}%
In each case, $2000$ events were generated for all three reaction systems
used in the analysis. The error bars were drawn to indicate fluctuations and
were obtained from the dispersion of $R_{i}$ time evolution \cite{chen05}.
First, it is interesting to note that with the in-medium NN cross sections
the strength of isospin transport $1-R_{i}$ decreases monotonically with the
decreasing value of $x$. With the free-space NN cross sections, however,
there appears to be a minimum at around $x=-1$. Moreover, this minimum is
the point closest to the experimental data. This allowed us to extract the
value of $K_{asy}(\rho _{0})=-550\pm 100$ MeV. With the in-medium NN cross
sections we can now further narrow down the $K_{asy}(\rho _{0})$ to be about
$-500\pm 50$ MeV. The latter is consistent with that extracted from studying
the isospin dependence of giant resonances of $^{112}$Sn to $^{124}$Sn
isotopes by Fujiwara et al at Osaka \cite{osaka}. Shown also in the figure
are the $\gamma $ values used in fitting the symmetry energy with $E_{\text{%
sym}}(\rho )=31.6(\rho /\rho _{0})^{\gamma }$. The results with the
in-medium NN cross sections constrain the $\gamma $ parameter to be between $%
0.69$ and $1.05$. The lower value is close to what is extracted from
studying giant resonances \cite{pie,colo}. The value of $\gamma =1.05$
extracted earlier using the free-space NN cross sections sets an upper
limit.
\begin{figure}[th]
\includegraphics[scale=1.4]{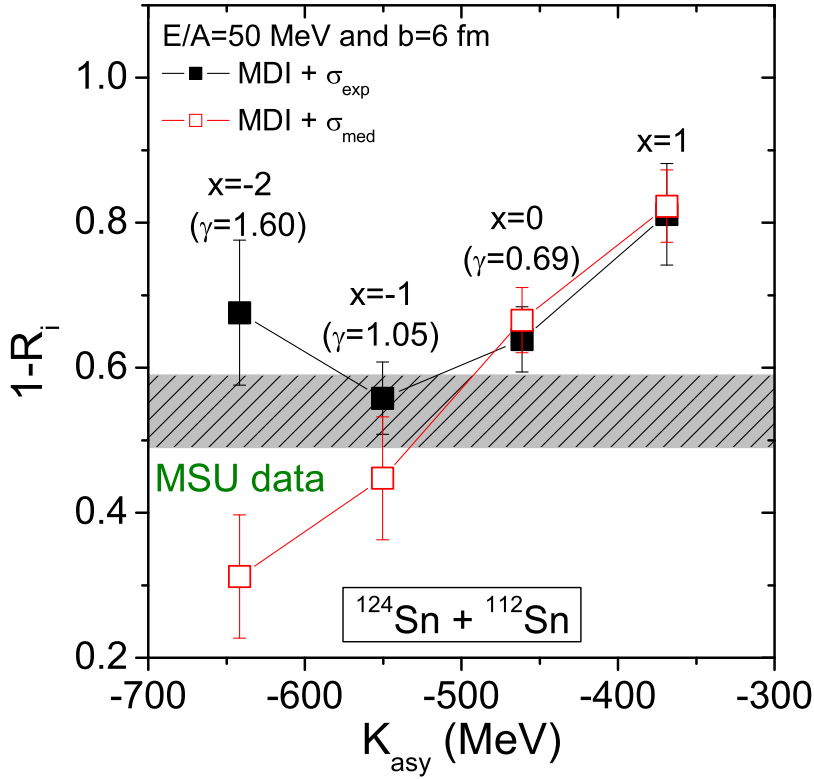}
\vspace{1.0 cm}
\caption{{\protect\small (Color online) The degree of isospin transport as a
function of $K_{\text{asy}}(\protect\rho _{0})$ with the free (filled
squares) and in-medium (open squares) nucleon-nucleon cross sections.}}
\label{figure8}
\end{figure}

It is seen that the difference in $1-R_{i}$ obtained with the free-space and
the in-medium NN cross sections is about the same with $x=1$ and $x=0$, but
then becomes especially large at $x=-1$ and $x=-2$. Why does the effect of
the in-medium NN cross sections increase with the decreasing $K_{asy}(\rho
_{0})$ or $x$ parameter? This question can be understood from considering
contributions from the symmetry potential and the np scatterings. As we have
mentioned in the introduction, both contributions to the isospin transport
depend on the np scattering cross section $\sigma _{np}$. Schematically, the
mean field contribution is proportional to the product of the isospin
asymmetric force $F_{np}$ and the inverse of the np scattering cross section
$\sigma _{np}$. While the collisional contribution is proportional to the $%
\sigma _{np}$. The overall effect of the in-medium NN cross sections on
isospin transport is a result of a complicated combination of both the mean
field and the NN scatterings. Generally speaking, the symmetry potential 
effects on the isospin transport will become weaker when the NN cross sections are larger
while the symmetry potential effects will show up more clearly if smaller NN
cross sections are used. This feature can be seen from Fig.\ 9.
\begin{figure}[th]
\includegraphics[scale=1.4]{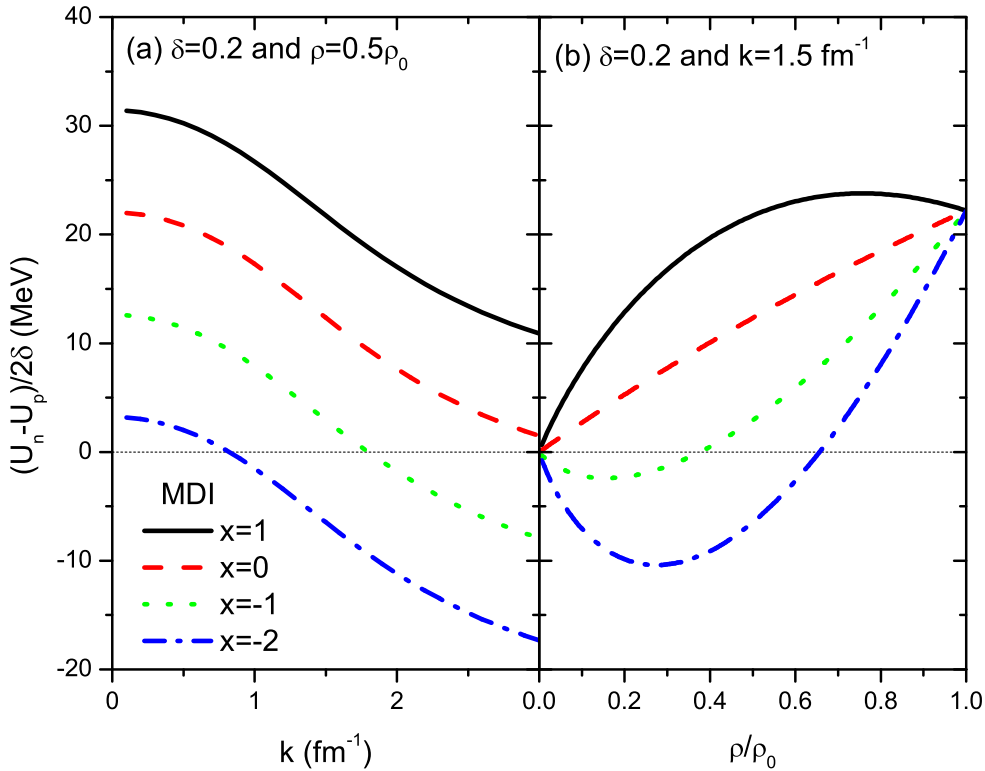}
\vspace{1.0 cm}
\caption{{\protect\small (Color online) Symmetry potential as a function of
momentum at selected densities (a), and density at selected momenta (b) with
the four $x$ parameters. Taken from Ref. \protect\cite{chen05}.}}
\label{figure9}
\end{figure}

For $x=1$ and $x=0$, the symmetry potential and its density slope, as shown
in Fig.\ 9, are large at low densities where the majority of net isospin transport
occurs. The $F_{np}$ factor makes the contribution due to the mean field
dominates over that due to the collisions. Therefore, the reduced in-medium $%
\sigma _{np}$ leads to about the same or a slightly higher isospin
transport. As the $x$ parameter decreases to $x=-1$ and $x=-2$, however, the
symmetry potential decreases and its density slope can be even negative at
low densities. Thus in these cases either the collisional contribution
dominates or the mean field contribution becomes negative. The reduced
in-medium np scattering cross section $\sigma _{np}$ leads then to a lower
isospin transport compared with the case with the free-space NN cross
sections.
\begin{figure}[tbp]
\includegraphics[scale=0.8,angle=-90]{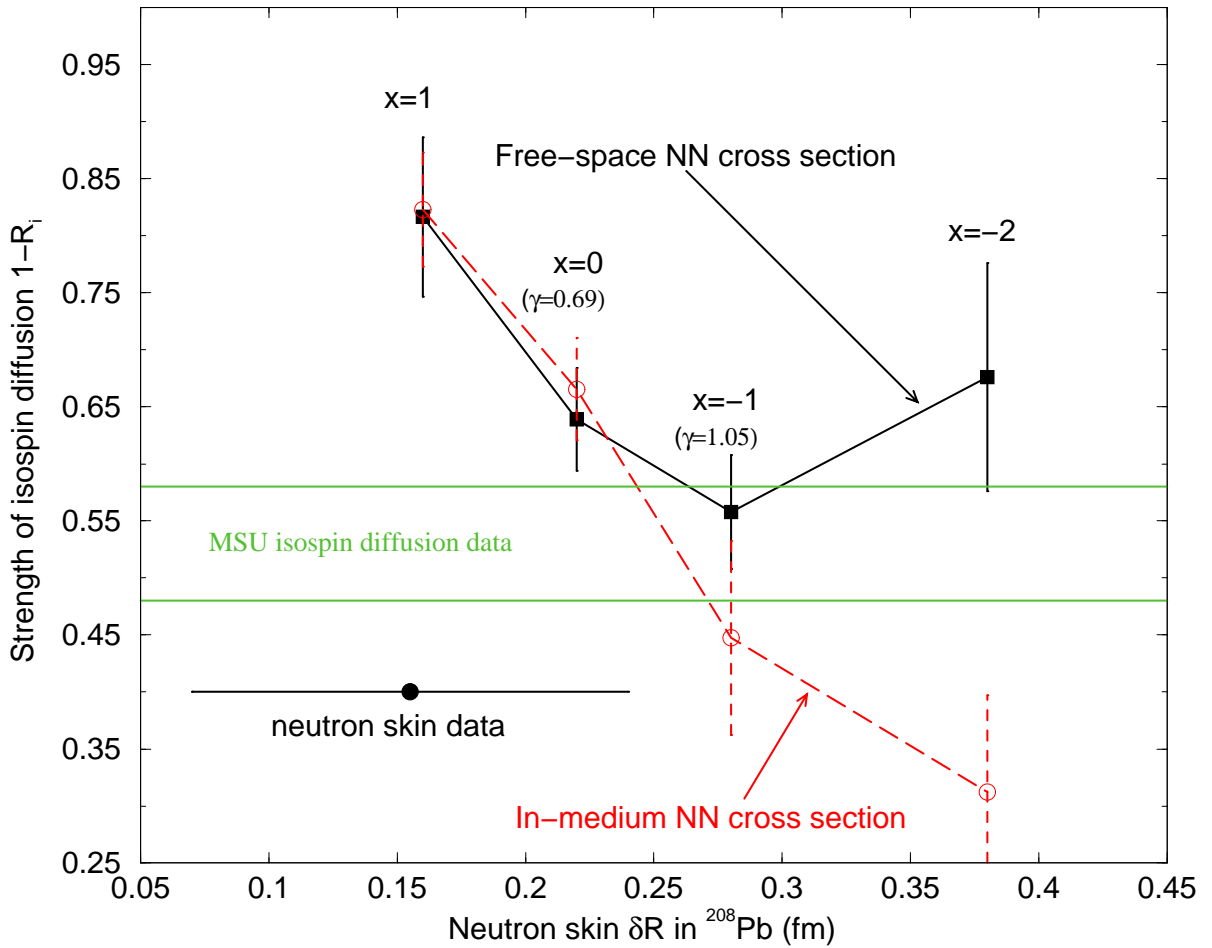}
\vspace{1.0 cm}
\caption{(Color online) The skin thickness in lead, $\protect\delta R$,
versus the isospin diffusion parameter $1-R_{i}$ for the four equations of
state in this work. The present acceptable experimental range for $\protect%
\delta R$ and the NSCL/MSU data on isospin diffusion are also shown.}
\label{figure10}
\end{figure}

\subsection{Correlation between the isospin diffusion and neutron-skin in $%
^{208}Pb$}

Both the neutron skin of heavy nuclei and the degree of isospin diffusion in
heavy-ion collisions at intermediate energies are sensitive to the symmetry
energy at sub-saturation densities. Since they both are determined by the
same underlying EOS, the correlation between these seemingly very different
observables provides a more stringent constraint on the symmetry energy than
their individual values \cite{steiner05}. In view of the revised
calculations of the isospin diffusion using the in-medium NN cross sections,
here we revisit the correlation between the isospin diffusion and the size
of neutron-skin in $^{208}$Pb. Shown in Fig.\ 10 is the isospin diffusion
parameter $1-R_{i}$ versus the skin thickness in lead, $\delta R$. The
presently acceptable experimental range for $\delta R$ \cite{nskin} and the
NSCL/MSU data on isospin diffusion are also shown. The $\delta R$, taken
from Ref. \cite{steiner05}, was calculated within the Hartree-Fock approach
using the same equations of state corresponding to the four $x$ parameters.
While the isospin diffusion data is between calculations with $x=0$ and $x=-1
$, the upper limit of the present n-skin measurements comes closer to the
prediction using $x=0$. On one hand, the present analysis of the isospin
diffusion data favors a neutron skin as large as about $0.25$ fm. On the
other hand, the present neutron-skin data favors $x=0$ and $x=1$. One can
thus conclude that a symmetry energy of $E_{\text{sym}}(\rho )\approx
31.6(\rho /\rho _{0})^{0.69}$ is currently most acceptable based on both the
neutron-skin and the isospin diffusion data. It is seen that more accurate
measurements of both kinds of experimental data, especially the
neutron-skin, are very desirable.

\section{Nucleon transverse flow as a probe of the in-medium NN cross
sections}

Our results above indicate clearly that the in-medium NN cross sections
affect significantly the extraction of symmetry energy from isospin
transport in heavy-ion reactions. Therefore, auxiliary measurements of other
observables have to be made to constrain the in-medium NN cross sections,
preferably from the same experiments. Fortunately, a number of observables
are known to be sensitive to the in-medium NN cross sections. These include
the quadruple moment $Q_{zz}$ of nucleon momentum distribution, the linear
momentum transfer (LMT) and the ratio of transverse to longitudinal energies
(ERAT), see, e.g., ref.\cite{pawel,cai05,reisdorf04,ldw}. It is also well
known that these observables are also sensitive to the EOS of symmetric
matter. Therefore, the extraction of the in-medium NN cross sections from
these observables relies closely on our knowledge about the EOS of symmetric
nuclear matter. Nevertheless, great progress has been made over the last
three decades in determining the EOS of symmetric nuclear matter. Using a
combination of several observables in heavy-ion reactions, such as the
elliptic flow and kaon production, see, e.g., Refs. \cite%
{reisdorf,science,fuchs}, and the analysis of giant resonances, the EOS of
symmetric nuclear matter has been severely constrained. It has now been
widely recognized that the major remaining uncertainty in further
constraining the EOS of symmetric nuclear matter is our poor knowledge about
the symmetry energy \cite{pie04,colo04,science}. In this work we thus
concentrate on the possibility of extracting the symmetry energy and the
in-medium NN cross sections simultaneously without considering the remaining
uncertainty in the EOS of symmetric matter.

In this work we choose to use the transverse flow to constrain the in-medium
NN cross sections. Its sensitivity to variations of the in-medium NN cross
sections is well known, especially around the balance energies, see, e.g.,
\cite{bert87,ogi,xu,mot,gar,klakow,li93,huang}. As an illustration, shown in
Fig.\ 11 is the transverse flow of all free nucleons identified as those
having local densities less than $\rho _{0}/8$ at freeze-out in the reaction
of $^{132}$Sn+$^{124}$Sn at an impact parameter of $5$ fm and a beam energy
of $400$ MeV/A (upper) and $50$ MeV/A (bottom) with the free-space and
in-medium NN cross sections, respectively. It is seen that at $50$ MeV/A the
transverse flow is much more sensitive to the in-medium NN cross sections
than at $400$ MeV/A. In fact, the direction of transverse flow is altered
from being positive to slightly negative by the reduced in-medium NN cross
sections. This observation is consistent with our expectation and results of
previous studies.
\begin{figure}[tbp]
\includegraphics[scale=1.,angle=-90]{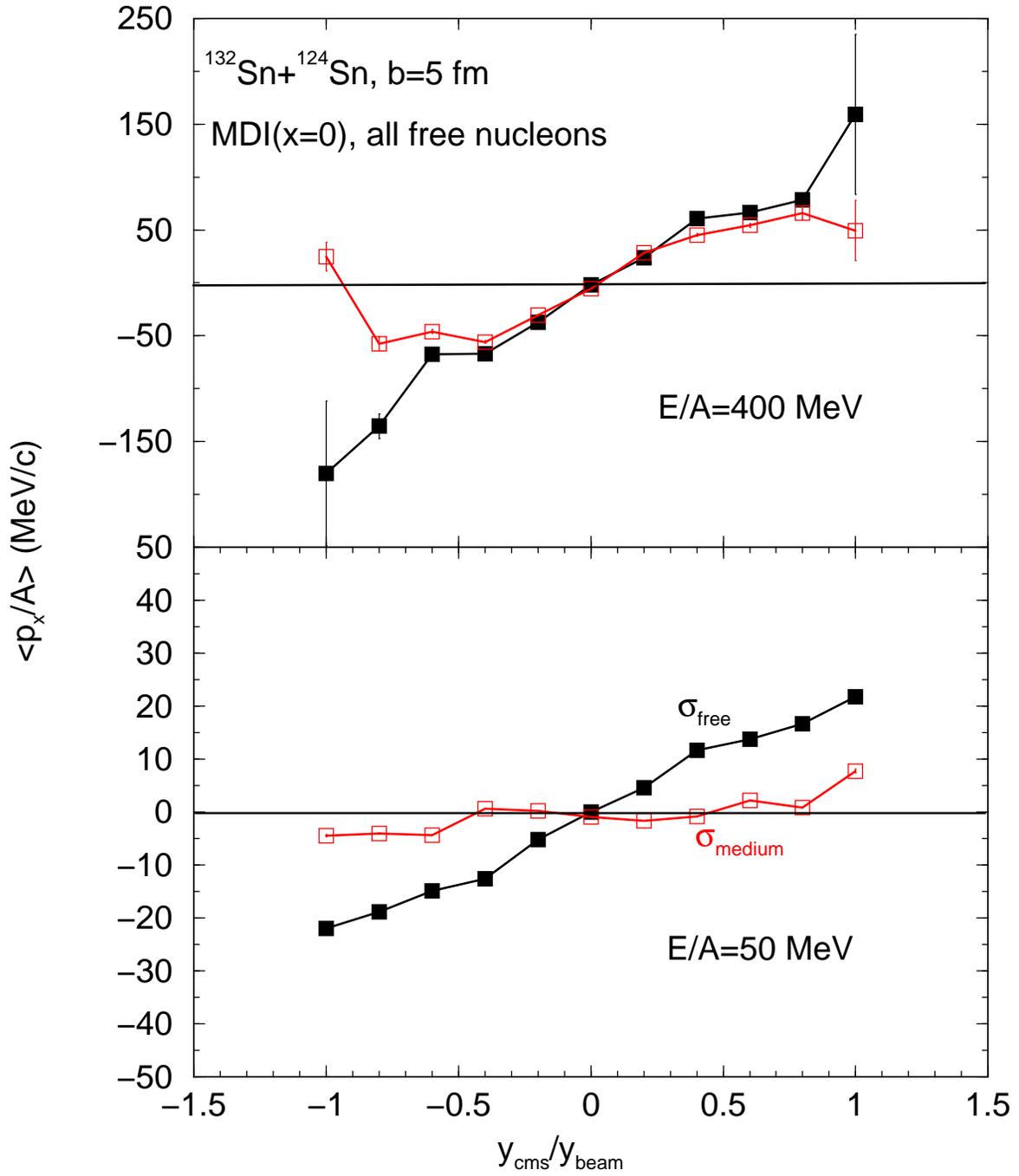}
\vspace{0.7 cm}
\caption{Nucleon transverse flow in the reaction of $^{132}$Sn+$^{124}$Sn at
an impact parameter of 5 fm and a beam energy of 400 MeV/A (upper) and 50
MeV/A (bottom).}
\label{figure11}
\end{figure}

\begin{figure}[tbp]
\includegraphics[scale=1.2]{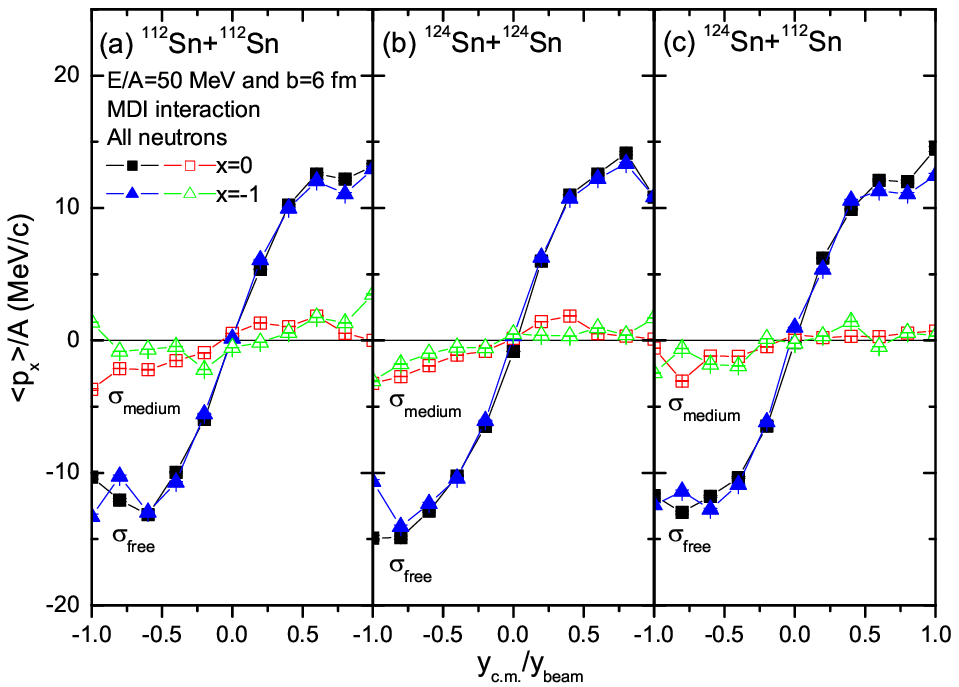}
\vspace{0.8 cm}
\caption{Neutron transverse flow in the reaction of $^{112}Sn+^{112}Sn$
(left), $^{124}Sn+^{124}Sn$ (middle) and $^{124}Sn+^{112}Sn$ (right) at an
impact parameter of 6 fm and a beam energy of 50 MeV/A.}
\label{figure12}
\end{figure}
\begin{figure}[tbp]
\includegraphics[scale=1.2]{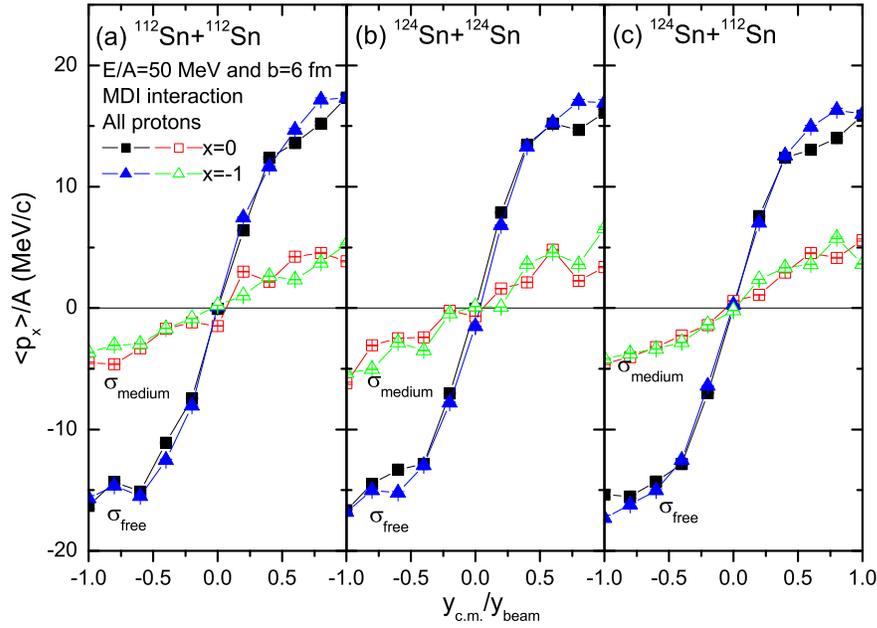}
\vspace{0.8 cm}
\caption{Proton transverse flow in the reaction of $^{112}Sn+^{112}Sn$
(left), $^{124}Sn+^{124}Sn$ (middle) and $^{124}Sn+^{112}Sn$ (right) at an
impact parameter of 6 fm and a beam energy of 50 MeV/A.}
\label{figure13}
\end{figure}

Shown in Figs.\ 12 and 13 are the transverse flow for all (free and bound)
neutrons and protons, respectively, with $x=0$ and $x=-1$ for the three
reactions used in studying the isospin transport. Free nucleons show the
same features but with poor statistics for the same number of events. There
is very little system size dependence among the three reactions considered.
It is seen that the transverse flow is much more sensitive to the in-medium
NN cross sections than to the symmetry energy parameter $x$. There is a weak
sensitivity to the variation of the $x$ parameter, especially with the
reduced in-medium NN cross sections for neutrons. Protons are affected by
both the repulsive Coulomb potential and the generally attractive symmetry
potential, while neutrons are only affected by the repulsive symmetry
potential, besides the same isoscalar potential acting on both neutrons and
protons. Neutrons thus appear to be more sensitive to the $x$ parameter.
Moreover, the Coulomb potential dominates over the symmetry potential,
leading to the slightly higher transverse flow for protons than for neutrons
as seen from comparing Fig. 12 with Fig. 13. It is also interesting to note
from comparing Fig.\ 12 with Fig.\ 13 that there is a clear indication of
higher (lower) transverse flow for neutrons (protons) with $x=0$ than that
with $x=-1$, especially around the projectile and target rapidities. This
feature is what one expects from considering the symmetry potentials shown
in Fig.\ 9. At sub-saturation densities the symmetry potential is stronger
with $x=0$ than that with $x=-1$. Therefore, the neutron (proton) transverse
flow is stronger (weaker) with $x=0$. This fine dependence on the symmetry
energy parameter $x$ can be studied in more detail by using the
neutron-proton differential flow \cite{dflow}. We find, however, that the
strength of the neutron-proton differential flow only changes by about $1$
to $2$ MeV/c by varying the $x$ parameter from $-1$ to $0$. It is much less
than the change due to the variation of the in-medium NN cross sections.
Based on these results, we therefore propose to measure the proton
transverse flow in order to constrain the in-medium NN cross sections.
\begin{figure}[tbp]
\includegraphics[scale=1.2]{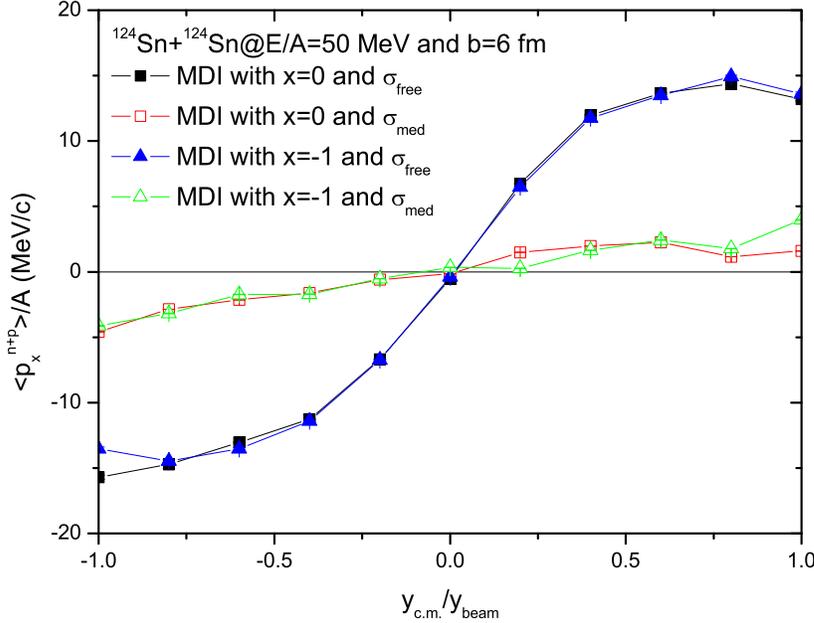}
\caption{Combined nucleon transverse flow in the reaction of $%
^{124}Sn+^{124}Sn$ at an impact parameter of 6 fm and a beam energy of 50
MeV/A.}
\label{figure14}
\end{figure}

To go one step further, taking into account the generally positive/negative
nature of the symmetry potentials for neutrons/protons, one may use the
combined transverse flow of neutrons and protons $<p_{X}^{n+p}/A>$ to better
determine the in-medium NN cross sections without much influence of the
symmetry potential. This, of course, requires measuring neutrons as
accurately as protons simultaneously. Shown in Fig. 14, is an example of the
combined nucleon flow for the $^{124}$Sn+$^{124}$Sn reactions. It is clearly
seen that the combined flow of all nucleons are much less affected by the
variation of the symmetry energy, making it an even better probe of the
in-medium NN cross sections.

\section{Summary}

In order to better understand the isospin dependence of the
in-medium nuclear effective interactions, we investigated effects
of the in-medium NN cross sections on isospin transport in
heavy-ion reactions within the transport model IBUU04. The
isospin-dependent in-medium NN cross sections consistent with the
nuclear mean field used in the transport model were evaluated by
using the scaling model according to the nucleon effective masses.
It is found that the NN cross sections in neutron-rich matter are
not only reduced compared with their values in free space, their
isospin dependence is also altered. Because of the positive and
growing neutron-proton effective mass splitting in more
neutron-rich matter for the effective interactions used in this
work, the splitting between the nn and pp cross sections increases
with the increasing isospin asymmetry of the medium.

The in-medium NN cross sections are found to influence significantly the
isospin transport and nucleon transverse flow in heavy-ion reactions at
intermediate energies. By using the free-space experimental NN cross
sections, a symmetry energy of $E_{\text{sym}}(\rho )\approx 31.6(\rho /\rho
_{0})^{1.1}$ was extracted from the MSU data on isospin transport. With the
in-medium NN cross sections, however, the symmetry energy of $E_{\text{sym}%
}(\rho )\approx 31.6(\rho /\rho _{0})^{0.69}$ was found most acceptable in
comparison with both the MSU isospin diffusion data and the presently
acceptable neutron-skin thickness in $^{208}$Pb. The isospin dependent part $%
K_{asy}(\rho _{0})$ of the isobaric imcompressibility of nuclear matter was
further narrowed down to $-500\pm 50$ MeV.

The possibility of determining simultaneously both the in-medium NN cross
sections and the symmetry energy corresponding to the same underlying
nuclear effective interactions was also studied. The proton transverse flow,
or even better the combined transverse flow of neutrons and protons all
together, can be used as an effective probe of the in-medium NN cross
sections without much hindrance from the uncertainties of the symmetry
energy. Our findings in this work demonstrated clearly the importance of
using the in-medium NN cross sections consistent with the momentum-dependent
nuclear mean field in transport model studies of heavy-ion reactions.

We would like to thank Dr. A.W. Steiner for helpful discussions. The work of
B.A. Li is supported in part by the US National Science Foundation under
Grant No. PHY-0354572, PHY0456890 and the NASA-Arkansas Space Grants
Consortium Award ASU15154. The work of L.W. Chen is supported in part by the
National Natural Science Foundation of China under Grant No. 10105008.

\end{document}